# Predicting the location of shear band initiation in a metallic glass

Z. Fan,[1] E. Ma[1] and M. L. Falk[1,2,*]

[1]Department of Materials Science and Engineering, Johns Hopkins University, Baltimore, Maryland 21218, USA
[2]Department of Physics and Astronomy, Department of Mechanical Engineering, and Hopkins Extreme Materials Institute, Johns Hopkins University, Baltimore, Maryland 21218, USA

*corresponding author.
mfalk@jhu.edu

*Abstract*

We report atomistic simulation results which indicate that the location of shear banding in a metallic glass (MG) can be ascertained with reasonably high accuracy solely from the undeformed static structure. Correlation is observed between the location of the initiation of shear bands in a simulated MG and the initial distribution of the density of fertile sites (DFS) for stress-driven shear transformations identified *a priori* based on a deep learning model devised in our recent work [Fan and Ma, Nat. Commun. **12**, 1506 (2021)]. In addition, we demonstrated that one can judge whether a glass is brittle or ductile solely based upon its initial DFS distribution. These validate that shear bands in MG arise from non-linear instabilities, and that the as-quenched glass structure contains inhomogeneities that influence these instabilities. This work also reveals an important subtlety regarding the non-deterministic nature of athermal quasistatic shear simulations.

*Introduction*.—The dominant flow and failure mode in metallic glasses (MGs) is shear banding, a plastic instability that localizes large amounts of shear strain within a narrow region when a MG is deformed [1-5], which is a *non-local* event. Understanding and controlling the shear banding behavior is of importance for enhancing the ductility of MGs, and thus, shear banding has been a research focus in the MG community over the past few decades [6-26]. However, owing to the disordered structure of MGs and the lack of well-defined topological defects like dislocations, it remains an open question whether one can identify *a priori* where shear bands will initiate solely based upon their as-quenched undeformed static structure. Also, it is not clear whether one can determine solely from a MG's initial static structure if the MG is brittle or ductile. A number of investigations have been undertaken seeking to apply machine learning (ML) methodologies to this problem. These have included simulations of amorphous polymer nanopillars [27] where surface properties and features are crucial to determining the mode of failure [28]. Our work builds on prior investigations using a linear support vector machine (SVM) [29] and graph neural networks (GNN) [30] to analyze bulk glass structure, although these prior works did not focus on predicting shear band formation per se. We also compare our convolutional neural network (CNN) based approach to physically derived indicators.

Previous research [15,16] has demonstrated that a shear band is formed by the coalescence of a series of adjacent shear transformation zones (STZs) [31], elementary plastic events that occur under externally applied stresses. Our conjecture is that one would be able to predict where shear bands initiate in MGs if one were to know which atoms would be involved in STZs under a specific loading condition *a priori* solely from the initial static structure. This would be consistent with models such as the STZ theory of amorphous plasticity wherein the shear banding instability arises from an underlying non-linear instability [32,33]. Over the years, much progress has been achieved in predicting elementary plastic events from initial structure in glasses [34,35]. Very recently, we demonstrated that even at 10% strain (around the yielding point) the atoms which will experience extremely large plastic rearrangement can be identified *a priori* with high accuracy solely from the static structure in the initial glassy sample before deformation [36]. This was achieved through the combination of a novel structural representation and a powerful deep learning method (convolutional neural network, CNN) [37]. In this letter, these atoms with high predicted plastic susceptibility (class probability > 0.5) via this method will be referred to as fertile sites for shear transformation, and we will demonstrate that shear bands prefer to initiate around regions with a high density of such fertile sites in simulated $Cu_{50}Zr_{50}$ MGs. Besides, we will also show whether a glass is brittle or ductile can be determined solely based upon its initial distribution of the density of fertile sites.

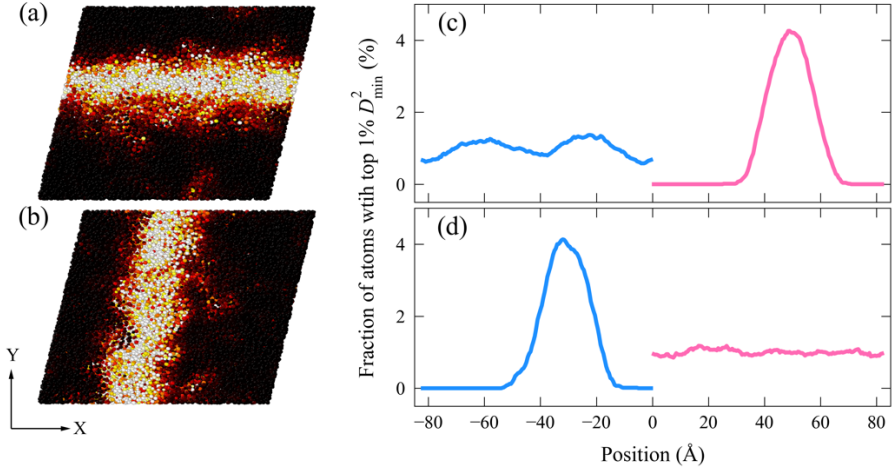

FIG. 1. Location of shear bands in a MG. (a) and (b) show color maps where bright (dark) atoms represent large (small) $D^2_{min}$ after athermal quasistatic shear to 20% strain is imposed on the same $Cu_{50}Zr_{50}$ MG sample under the same loading condition but from simulation runs in which the atom sequence is re-ordered. Plots (c) and (d) show the distribution of the fraction of atoms in the top 1% of the $D^2_{min}$ distribution in (a) and (b), respectively. The location of the distribution along the X axis (blue curve) was multiplied by -1 in order to differentiate it from that of the distribution along Y axis (pink curve).

*Initiation sites of shear bands in a MG.*—A recent study by Golkia, et al. [38] revealed that for identical Lennard-Jones glass models that were quenched instantaneously from supercooled liquids and then annealed, small changes in the deformation protocol, such as the use of different initial random numbers for the thermostat, changed the flow patterns from vertical to horizontal bands, and could result in horizontal bands arising at different locations. This occurred even though the shear deformation was conducted at very low temperature. From this the authors concluded that the initial undeformed glass sample does not determine whether and where shear bands will form. Rather, they assert that the formation of shear band is entirely stochastic. We have also studied deformation under athermal quasi-static shear (AQS) [39,40], a methodology that is usually considered to be deterministic since deformation is imposed in small shear increments followed by energy minimization operations. In our prior study we discovered that the AQS result varies if the atom sequence in the initial configuration file is reordered while keeping all atom positions and deformation conditions the same. Furthermore, the larger the strain, the larger the difference between simulations with different atom sequences [36]. This unexpected stochasticity appears to arise from the effect of numerical noise on the conjugate gradient minimization within the high-dimensional energy landscape of the glass, although other minimization algorithms were also tried and the similar stochasticity was observed. In this study we examine the degree to which shear banding in a relatively slowly quenched glass is stochastic with respect to location and orientation. We wish to ascertain whether the result of Golkia, et al. [38] is specific to the way their systems were prepared and simulated.

To this end, we simulate the identical $Cu_{50}Zr_{50}$ MG system consisting of 32,000 atoms in a cubic simulation cell with an edge length of approximately 82 Å. This initial condition was prepared by a quench at an effective cooling rate of $10^{10}$ K/s (other details regarding the sample preparation are the same as that in Ref. [36]). The system was sheared 100 times under exactly the same loading conditions using the AQS method on the same computer. The only difference among the 100 deformation simulations is the atom sequence in the file of the initial undeformed configuration. In this way we hope to determine the reproducibility of the shear band location and orientation.

Fig. 1(a) and (b) show the field of deviation from affine displacement ($D^2_{min}$) [31] at 20 % strain for two typical simulations among the 100 performed (Both the calculation and visualization of $D^2_{min}$ field were performed using the OVITO package [41]). As can been seen from the two $D^2_{min}$ fields, a horizontal shear band (the bright region in Fig. 1(a), parallel to X axis along the shear direction) formed in one simulation while a vertical shear band (the bright region in Fig. 1(b), parallel to the Y axis perpendicular to the shear direction) formed in another simulation. This demonstrates that the location of a shear band is not fully reproducible for the same MG model even under AQS simulation. However, one cannot reach the conclusion that the formation of shear bands is fully stochastic with respect to location for the slowly quenched glass subjected to AQS by considering select cases. The entire data bank generated of 100 simulations must be examined in a proper statistical analysis, as will be done in the next section.

In order to determine the center position of each shear band, since only one shear band occurs in each of the simulated samples, we calculate the locations of the atoms in the top 1% ($=f_{top}$) of the $D^2_{min}$ distribution at 20% shear strain. We will refer to these atoms as the "highly deformed atoms." As the shear bands have two possible orientations (horizontal or vertical), the distribution along both X and Y axes are examined. We calculate the fraction of highly deformed atoms within parallelepipeds that have a width along the X axis $\omega_{SB} = 19$ Å. The length along all other dimensions are equal to that of the simulation box. We will denote the middle point of each parallelepiped aligned with the X axis as $x$. Similarly, we calculate the fraction of highly deformed atoms along the Y axis at the locations denoted $y$.

To differentiate the locations along the X and Y axes, the coordinate along the X axis was multiplied by -1 in all figures. Thus, for a horizontal shear band, the peak position of the distribution should be positive while the peak position is negative for a vertical shear band. Fig. 1(c) and (d) display the distribution of the fraction of highly deformed atoms corresponding to Fig. 1(a) and (b), respectively. Systematically varying the values for both $f_{top}$ and $\omega_{SB}$ does not result in significant change in the center position, see Figs. S1 and S2 of the Supplemental Material.

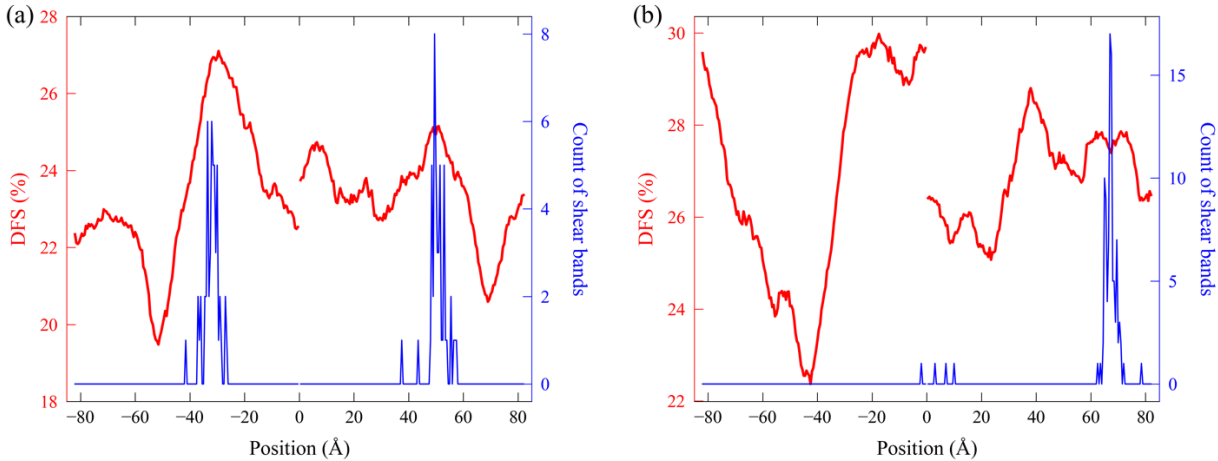

FIG. 2. Connection between initial distribution of density of fertile sites and final location of a shear band in a $Cu_{50}Zr_{50}$ MG. (a) and (b) are for two independent samples, respectively. In each plot, the red curve (left vertical axis) displays the initial distribution of CNN-predicted density of fertile sites (DFS), and the blue curve (right vertical axis) shows the number of shear bands initiated at each position obtained by shearing the same MG configuration 100 times under exactly the same loading condition but with different atom sequences.

*Predicting shear band locations from initial density of fertile sites*.— If the location of the initiation of a shear band is determined by the glass structure, deforming the same glassy sample multiple times under identical loading condition should result in repeated shear band initiation at highly correlated locations. To test whether this is the case, we count the number of shear bands initiated at each position for the 100 shearing simulations conducted on the same $Cu_{50}Zr_{50}$ MG sample under identical loading condition using the AQS method. As previously noted, before each simulation run the atom sequence was reshuffled. For the sample shown in Fig. 2(a) shear bands are observed to initiate predominately around two regions, and 48 out of the 100 shear bands are horizontal while the remaining 52 shear bands are vertical. We also sheared another independent $Cu_{50}Zr_{50}$ MG sample with identical processing history 100 times under the same protocol and found that for this initial condition 99 out of the 100 shear bands are horizontal and most of them initiated in the same region, while only 1 results in a vertical shear band, as shown in the blue curve (right vertical axis) in Fig. 2(b). These observations indicate that the shear band location is strongly determined by the initial static structure in these relatively gradually quenched glasses, and that the strain localization process is actually not purely stochastic. Based on the reasoning presented in the *Introduction* section, we anticipate that the density of fertile sites (DFS) predicted by the CNN model introduced in Ref. [36] should be higher around the peak positions of the blue curves shown in Fig. 2 relative to other regions.

To test this hypothesis, we took an approach similar to that used to determine the center position of the shear bands. We calculated the distribution of the DFS in the two independent samples before deformation, i.e., fraction of atoms with class probability > 0.5 in each of the parallelepiped regions. Here we again chose a value of 19 Å for $\omega_{DFS}$, the width of the parallelepiped along the X (or Y) dimension. This value of $\omega_{DFS}$ optimizes the correlation between the initial distribution of the DFS and the location of shear band initiations as shown in Fig. S3 of the Supplemental Material. The red curve (left vertical axis) in each panel of Fig. 2 displays the initial distribution of the DFS in the two independent samples, respectively. Indeed, the peak positions of the distribution of the DFS generally overlap with the peaks of the corresponding blue curve in Fig. 2, which is consistent with our expectation and suggests there is correlation between the initial static structure, i.e. the DFS predicted with our CNN model, and the sites of shear band initiation in MGs.

We note that the overlap in Fig. 2(a) is much more obvious than that in Fig. 2(b) where the predominant location of shear banding is near the second highest peak for horizontal bands, which is significantly smaller than the peak for vertical bands. This suggests that the structural order revealed by the DFS is likely only one of a number of factors that contributes to determine the location of the shear band.

We also note that, as evident from Fig. 2, sometimes no shear band coincides with the peak of DFS. This implies that the correlation revealed here is only statistically meaningful. In other words, while we can determine that some regions have higher propensity for shear band initiation, it remains difficult to predict the exact position of shear band initiation in a MG solely based on the initial static structure. This is because shear band initiation in MGs is a stochastic process, and this is true even in simulations of athermal deformation due to the numerical noise. Similar randomness is also expected to exist in real MG samples primarily due to thermal fluctuation. How best to characterize this stochasticity warrants further investigation.

*Figure of merit for the correlation.*—In the preceding section, we showed that the shear band location is potentially predictable solely based on the initial DFS distribution in a MG. In this section, we evaluate the predictive power of this approach by surveying 480 cases. To do this we prepared 20 $Cu_{50}Zr_{50}$ MG samples (not used in the training of our CNN model) and then sheared each of the 20 samples to 20% strain with the AQS method along 24 loading orientations using the strategy introduced in Ref. [36]. For each sample in each loading orientation, we first map out the CNN-predicted DFS along both the X and Y dimensions. Inspired by the methodology introduced in Ref. [42], we then define a parameter $\gamma$ as a figure of merit, the fraction of locations at which the DFS is lower than or equal to, that at the location of the shear band. If the shear band preferentially initiates in the region having the highest DFS, $\gamma$ would approach 1.0 for all cases, and the cumulative distribution function $C(\gamma)$ of $\gamma$ would be a step function: $C(\gamma) = 1$ for $\gamma = 1$ and 0 otherwise. At the other extreme, if there is no correlation between the DFS and the location of shear band initiation, the value of $\gamma$ should be stochastic and $C(\gamma)$ should be close to a straight line, i.e. the dashed diagonal line in Fig. 3(a). The resulting $C(\gamma)$ data are denoted by the red curve shown in Fig. 3(a), revealing a positive correlation.

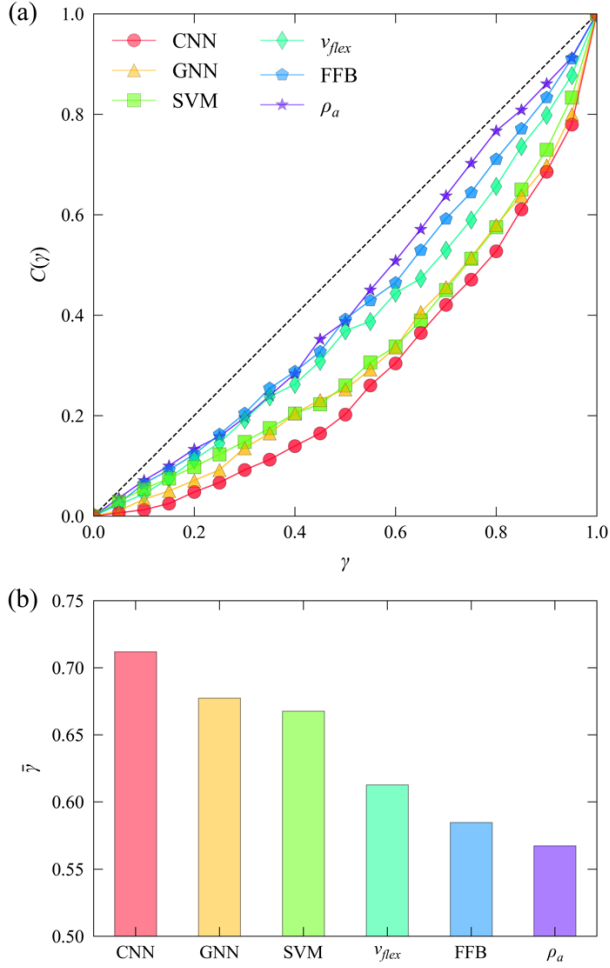

FIG. 3. Quantifying and comparing the predictive power of different methods in regards to shear band formation. (a) shows the cumulative distribution function $C(\gamma)$ of $\gamma$ (see text) for the predictions based on six different methods: CNN, graph neural network (GNN), linear support vector machine (SVM), flexibility volume ($v_{flex}$), five-fold bonds (FFB) and volume density of atoms ($\rho_a$). The diagonal dashed line represents random distribution with no correlation. A perfect prediction would entail $C(\gamma) = 1$ for $\gamma =1$ and 0 otherwise. (b) shows $\bar{\gamma}$ (arithmetic mean of $\gamma$) for the prediction based on the six methods. $\bar{\gamma} = 0.5$ means no predictive power and $\bar{\gamma} = 1.0$ perfect predictive power.

To show the advantage of our CNN-predicted DFS, we compare with predictions based on other data-driven models such as that derived from a graph neural network (GNN) [30], a linear support vector machine (SVM) [29], and three methods based on physical parameters, flexibility volume ($v_{flex}$) [43], fraction of five-fold bonds (FFB) [44] and the atomic density in the local region ($\rho_a$, an analog for free volume [45,46]). These metrics were all previously used in attempts to predict stress-driven shear transformations. Specifically, the fertile sites are defined as those atoms with GNN-predicted class probability > 0.5, or SVM-predicted distance to the separation boundary > 0, or within the top 20% of the values of $v_{flex}$ (we tried different thresholds for $v_{flex}$ but did not see obvious difference; average $v_{flex}$ within each band was tried as well and the resultant predictive power is slightly lower). For FFB and $\rho_a$, a lower value is expected to be more favorable for shear band initiation; we therefore modify $\gamma$ as the fraction of locations at which the density of FFBs or $\rho_a$ is *higher* than or equal to that in the region where a shear band is initiated. As seen from Fig. 3(a), the prediction based on our CNN-predicted DFS is superior to those from all the other five methods.

We also can use $\bar{\gamma}$ (the arithmetic mean of $\gamma$) to quantify the predictive power for these methods as $\bar{\gamma} = 1.0$ corresponds to perfect prediction and $\bar{\gamma} = 0.5$ would imply no correlation. As seen from the bar chart shown in Fig. 3(b), the $\bar{\gamma}$ of 0.712 for our CNN-based prediction is the highest. The advantage of our CNN over other ML methods is due to: i) the completeness of our new structural representation, the spatial density map (SDM); ii) the stronger learning capability of the CNN models, compared to the SVM models and GNN models used in previous studies; and iii) the anisotropy of local mechanical response of glasses has been taken into account in the CNN framework but ignored in all previous ML methods, see more discussion in Ref. [36]. Other ML methods such as conventional neural networks and the one proposed in Ref. [47] have been found to be comparable to the SVM method [35].

The structural representation and training procedure for all deep or machine learning methods in the current work also follow the procedures detailed in Ref. [36]. In this work, the fertile sites were predicted using the CNN, GNN or SVM model, each of which was trained at a strain of 10% with $f_{thres}$=5.0% based on the results shown in Figs. S4 and S5 of the Supplemental Material.

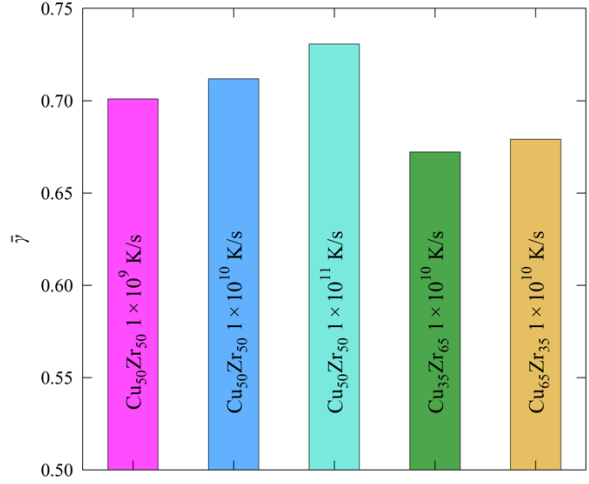

FIG. 4. The $\bar{\gamma}$ achieved with the CNN model on glasses with different processing histories or at different compositions (denoted on each bar).

The transferability of CNN models to different processing histories or compositions within the same alloy system has been demonstrated when predicting elementary plastic events in other recent work [36]. Therefore, success at one composition and processing history give us reason to expect that the CNN-based predicted DFS distribution can be used to predict the location of shear band initiation of glasses with different processing histories or at different compositions within the same alloy system. We have undertaken a number of simulations to confirm this, as shown in Fig. 4. We note that the $\bar{\gamma}$ achieved on a more quickly quenched glass is higher, which may be because deformation is more evenly distributed in such as system.

*Brittle versus ductile*.—The results above demonstrate that shear bands prefer to initiate at locations with higher CNN-based predicted DFS. This implies that plastic flow will not localize into a narrow region but will instead distribute over the entire sample, i.e., its deformation would be ductile, if the difference of DFS is small within a glass. Thus, it may be possible for us to judge if a glass would be brittle or ductile, which is another problem of interest to the community [48], based on the degree of fluctuation of the initial DFS distribution.

To verify this conjecture, we slowly reduce the temperature of a $Cu_{50}Zr_{50}$ liquid model containing 31,250 atoms from 2,500 K to various temperatures ($T_q$) at a constant rate of $1\times10^{10}$ K/s and then instantly quench them to 0 K via energy minimization. Next, we calculate the CNN-based predicted DFS distribution of these glasses with different $T_q$, and use the difference between the largest and smallest value of DFS distribution to denote the degree of fluctuation of the initial DFS distribution (Δ). As seen from Fig. 5, it is easy to find a critical Δ value (6.0% denoted by the red dashed line in Fig. 5) that separates these glasses into low and high $T_q$ regimes. By shearing these glasses via the AQS method, we confirmed that a shear band was formed in almost all glasses with Δ > 6% (The only exception is the glass with $T_q$ = 850 K) and no shear band appeared in glasses with Δ < 6%. Fig. S6 in the Supplemental Material shows the projection of the $D^2_{min}$ field on the XY plane for all of these glasses after strained to 20%. Two typical snapshots are shown as the insets in Fig. 5. These results confirmed that one can determine whether a glass is brittle or ductile based on its initial static structural information, i.e., the degree of fluctuation of CNN-based predicted DFS distribution.

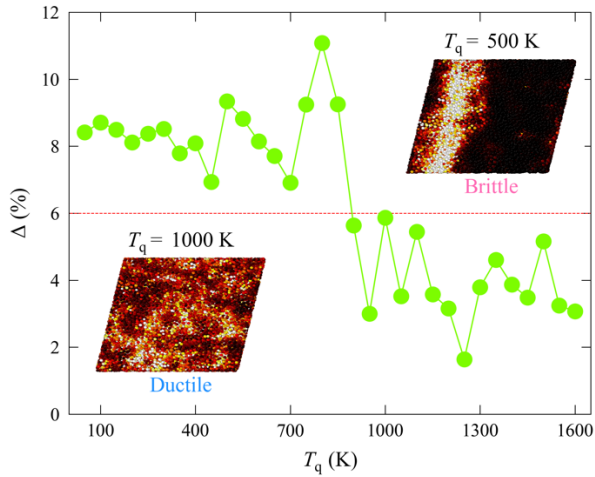

FIG. 5. Brittle versus ductile. The degree of fluctuation of CNN-based predicted DFS distribution (Δ) as a function of $T_q$. The red dashed line denotes the critical value of Δ we suggested to separate brittle and ductile glasses. The right upper (left lower) inset shows the $D^2_{min}$ filed of a brittle (ductile) glass. Bright and dark color correspond to large and small $D^2_{min}$, respectively.

*Conclusions.*—In summary, we sheared the same simulated (gradually quenched) MG sample under the same loading condition using the AQS method and varying atom sequence in the initial configuration. We found that shear bands initiate along one of the two maximum-shear planes, but, in doing so, prefer to initiate at locations with higher CNN-based predicted DFS, although this correlation is not perfect. We defined a figure of merit to quantify the correlation strength between CNN-based predicted DFS and the shear band initiation in the MG samples. We find that this indicator is superior in making this prediction when compared to a number of other structural indicators suggested previously. To the best of our knowledge, this is the first evidence of correlation between the initial static structure and shear band locations in simulated MGs. We also demonstrated whether a glass is brittle or ductile can be ascertained solely based upon its initial DFS distribution. The demonstrated correlation may be useful for establishing robust physical models of plastic behavior in amorphous solids. It also provides some optimism for controlling shear banding behavior and thus optimizing mechanical properties of MGs via the tuning of their initial structural state before deformation. At the same time, this work points to the need to better understand the stochasticity of the strain localization process as well as factors other than DFS that may play a role in determining how and where these bands develop.

*Acknowledgements*

The work is supported at JHU by U. S. Department of Energy (DOE) DOE-BES-DMSE under grant DE-FG02-03ER46056. E.M. left the project in September 2020. This research used the CPU resources of the National Energy Research Scientific Computing Center (NERSC), a DOE Office of Science User Facility supported by the Office of Science of the U.S. Department of Energy under Contract No. DE-AC02-05CH11231, and the Maryland Advanced Research Computing Center (MARCC). The authors also acknowledge the Texas Advanced Computing Center (TACC) at the University of Texas at Austin for providing GPU resources that have contributed to the research results reported within this paper.